\documentclass[prl,showpacs]{revtex4}

\input epsf

\begin{document}

\title{Measuring Neutron Star Radii with Gravitational Wave Detectors}
\author{Joshua A.\ Faber, Philippe Grandcl\'{e}ment, Frederic A.\ Rasio}
\affiliation{Department of Physics and Astronomy, 
 Northwestern University, Evanston, IL 60208}

\author{Keisuke Taniguchi}
\affiliation{Department of Earth Science and Astronomy,
Graduate School of Arts and Sciences,
University of Tokyo,
Komaba, Meguro, Tokyo 153-8902, Japan}

\date{\today}

\begin{abstract}
Coalescing binary neutron stars (NS) are expected to be an important
source of gravitational waves (GW) detectable by laser
interferometers.  We present here
a simple method for determining the compactness ratio $M/R$ of NS
based on
the observed deviation of the GW energy spectrum from point-mass behavior
at the end of an inspiral event. Our method is based on the
properties of quasi-equilibrium binary NS sequences and does not
require the computation of the full GW signal $h(t)$. Combined with
the measurement of the NS masses from the GW signal during
inspiral, the determination of $M/R$ will allow very strong constraints
to be placed on the equation of state of nuclear matter at high densities.
\end{abstract}
\maketitle

Coalescing compact binaries containing two neutron stars (NS) 
are among the most important sources of gravitational waves (GW) for
LIGO \cite{LIGO}, VIRGO \cite{VIRGO}, and other laser interferometers.  
Should the inspiral
of such a binary be detected, the frequency evolution of the GW signal
will immediately yield the system's ``chirp mass'' 
$M_{\rm ch}\equiv \mu^{3/5}{\cal M}^{2/5}$, 
where $\mu$ and ${\cal M}$ are the reduced
and total mass of the binary, respectively. Higher-order post-Newtonian
effects on the phase evolution of the signal also 
allow for the determination of the reduced mass $\mu$, and
thus the individual masses $M_1$ and $M_2$ of the two
NS \cite{CF}. The determination of the NS radii in 
addition to their masses would yield important
information about the equation of state (EOS) at nuclear densities,
and could even indicate the presence of more exotic phases, such as
strange quark matter instead of ordinary nuclear matter
\cite{Strange}.  The GW signal of a coalescing binary could yield
such information but this is limited in two different ways. 
During the slow inspiral phase at large
separations, i.e., low frequencies ($f\ll 1\,$kHz), the stars
behave like point masses, and finite-size effects are not
expected to leave any signature in the GW signal \cite{CF,Koch,Lai}.
During the final hydrodynamic merger, characteristic frequencies
of GW emission could yield important information about the fluid EOS
\cite{Zhu,FR,Oech2}, but these frequencies are expected to lie well
above the photon shot noise limit of current interferometers ($f\gtrsim 1\,$kHz).  
Thus, it is only during the last few orbits of the inspiral, just prior
to merger, that we can hope to see the imprint of the NS radii on a
measurable GW signal (with $f\lesssim 1\,$kHz).

Several groups have studied this terminal phase of inspiral
by constructing quasi-equilibrium sequences of close NS
binaries in the conformal flatness approximation of general 
relativity (GR) \cite{Baumg,Marro,Uryu00,Gourg01}.
In this approximation, it is assumed that the binary 
system evolves along a sequence of appropriately constructed 
equilibrium states with
decreasing binary separation as energy is radiated away in GW.
From the binary equilibrium energy curve
$E_{\rm eq}(r)$, which gives the total system energy as a function of
binary separation $r$, and the GW luminosity $L_{\rm GW}$,
one can derive the radial infall rate as 
$v_r=L_{\rm GW} (dE_{\rm eq}/dr)^{-1}$.
With $v_r=dr/dt$ known, this provides the time evolution along the 
equilibrium sequence and the GW signal $h(t)$. 
This approach should remain accurate as
long as the radial infall timescale $r/v_r$ is longer than the dynamical
timescale of the system, i.e., until the point where dynamical instability 
sets in, and the two stars plunge inward rapidly and merge.

Unfortunately, calculating the correct GW luminosity for a given matter 
configuration in GR is an extremely difficult task.
Different approaches have required either
time integration of the full non-linear equations of GR \cite{Duez},
or the solution of a complicated wave equation for terms representing
the spherical harmonic expansion of the GW metric perturbation \cite{SU}.  
The great complexity of these approaches is in stark contrast with 
the simplicity of the quasi-equilibrium approximation.
However, we point out here that the GW {\em energy spectrum\/}
$dE_{\rm GW}/df$ 
can be calculated directly and very simply from the equilibrium energy curve, 
independent of any knowledge about the GW luminosity.
Indeed, by definition of the quasi-equilibrium approximation, the energy 
decrease $-dE_{\rm eq}$ between two neighboring binary configurations
along the sequence is equal to the energy $dE_{\rm GW}$ radiated away
as the wave frequency sweeps up by $df$, where the GW frequency is twice
the orbital frequency, $f=2f_{\rm orb}$.
Thus, one should simply compute the total energy $E_{\rm eq}$ as a function
of frequency $f$ along the equilibrium sequence, and the GW energy spectrum
is then given by the derivative $dE_{\rm GW}/df = - dE_{\rm eq}/df$.
As a trivial example, consider the inspiral of two point masses in
the Newtonian limit, where we have $E_{\rm eq} \propto -r^{-1}$ and
$f\propto r^{-3/2}$.  It follows that $E_{\rm eq}
\propto -f^{2/3}$ and thus $dE_{\rm GW}/df \propto f^{-1/3}$, a 
well-known result.
In addition to the assumptions underlying the quasi-equilibrium approximation,
the validity of this simple approach relies on the additional assumption
that the GW emission during the later merger phase has no effect on
the energy spectrum at lower frequencies. Indeed, this has been
demonstrated in numerical hydrodynamic calculations of binary mergers
\cite{Zhu,FR}, which show a clear separation between the inspiral and
merger components of the emission in frequency space.  

We have investigated the properties of the GW emission during the final
phase of binary NS inspiral using new, highly accurate equilibrium 
sequences calculated with the Meudon code \cite{Lorene}.
The formalism is based on the conformally flat approximation,
which is expected to yield accurate equilibrium matter configurations 
for this phase \cite{Baumg,Marro,Uryu00,Gourg01}.  The resulting five
non-linear, coupled elliptic equations are solved using a multi-domain 
spectral method \cite{Bonaz,Grand01}. This approach has already been used
successfully in various astrophysical applications
\cite{Gourg01,Gourg99,Grand02}.  
Typically, the computed fields satisfy the constraints of full GR to within
$\sim 1\%$ \cite{MMW}.
The code has been improved recently, especially with regard to the
treatment of the external compactified domain, and the numerical error
in computing equilibrium configurations, measured in terms of how well 
the virial theorem is satisfied, is of order one part in $10^{5}$. 
The equilibrium sequences presented here are the natural extension 
of the work already published in \cite{Gourg01,TaniII}.
They will be discussed in more detail in \cite{TaniIV}. 
Here we only show the variation of the ADM mass of the system
(total binary mass-energy ${\cal M}\equiv c^{-2} E_{\rm eq}$)
and the GW frequency (twice the orbital frequency), which are sufficient 
for our purposes to determine the GW energy spectrum.

The equilibrium configurations have been calculated for irrotational
binaries, i.e., assuming that the fluid has zero vorticity in the inertial 
frame. Indeed, the NS should be spinning slowly at large separations and the
viscosity of NS matter is too small for tidal spin-up to become
significant on the coalescence timescale \cite{Koch,BC}. 
Based on the current set of well-measured NS masses
in relativistic binary radio pulsars, it is expected that all NS in
coalescing binaries will have masses $M\simeq 1.35 M_\odot$ \cite{TC}.
Hence, for simplicity, we consider only equal-mass binaries where
$M_1=M_2=1.35 M_\odot$. Also for simplicity,
we model the NS EOS with a simple polytropic form
$P= K \rho^\Gamma$, where $P$ is the pressure and $\rho$
the rest-mass density. The constant $K$ represents the
overall compressibility of the matter and largely sets the value of
the stellar radius for a given mass, while the adiabatic exponent
$\Gamma$ measures the stiffness of the EOS and affects the degree
of central concentration of the NS interior. Based on our experience
with hydrodynamic calculations \cite{FR}, we expect that the GW energy
spectrum just prior to merger is determined primarily by the NS radius 
through $K$, with relatively little sensitivity to $\Gamma$. For this 
reason, 
in this initial study, we allow $K$, and therefore also the stellar radius 
$R$, to vary for different NS models, but we set $\Gamma=2$ for all models,
as this value fits well most published NS EOS (see,
e.g., \cite{Lattimer} and references therein).  
Specifically, we consider NS models with compactness
ratios $M/R=\,$0.12, 0.14, 0.16, and 0.18 (setting $G=c=1$), where $M$ is the ADM 
(gravitational) mass measured by an observer at infinity for a single
isolated NS, and $R$ is the circumferential radius of the NS.
For $M=1.35\,M_\odot$, the corresponding radii are 
$R=\,$16.6, 14.2, 12.4, and 11.1 km, respectively, spanning the range 
of values for NS radii calculated from various physical EOS.  
Note, however,
that our results are unchanged under the rescaling given by $R'=\kappa R$,
$M'=\kappa M$, $f'=f/\kappa$, for any constant $\kappa$. 

\begin{figure}
\centering \leavevmode \epsfxsize=6in \epsfbox{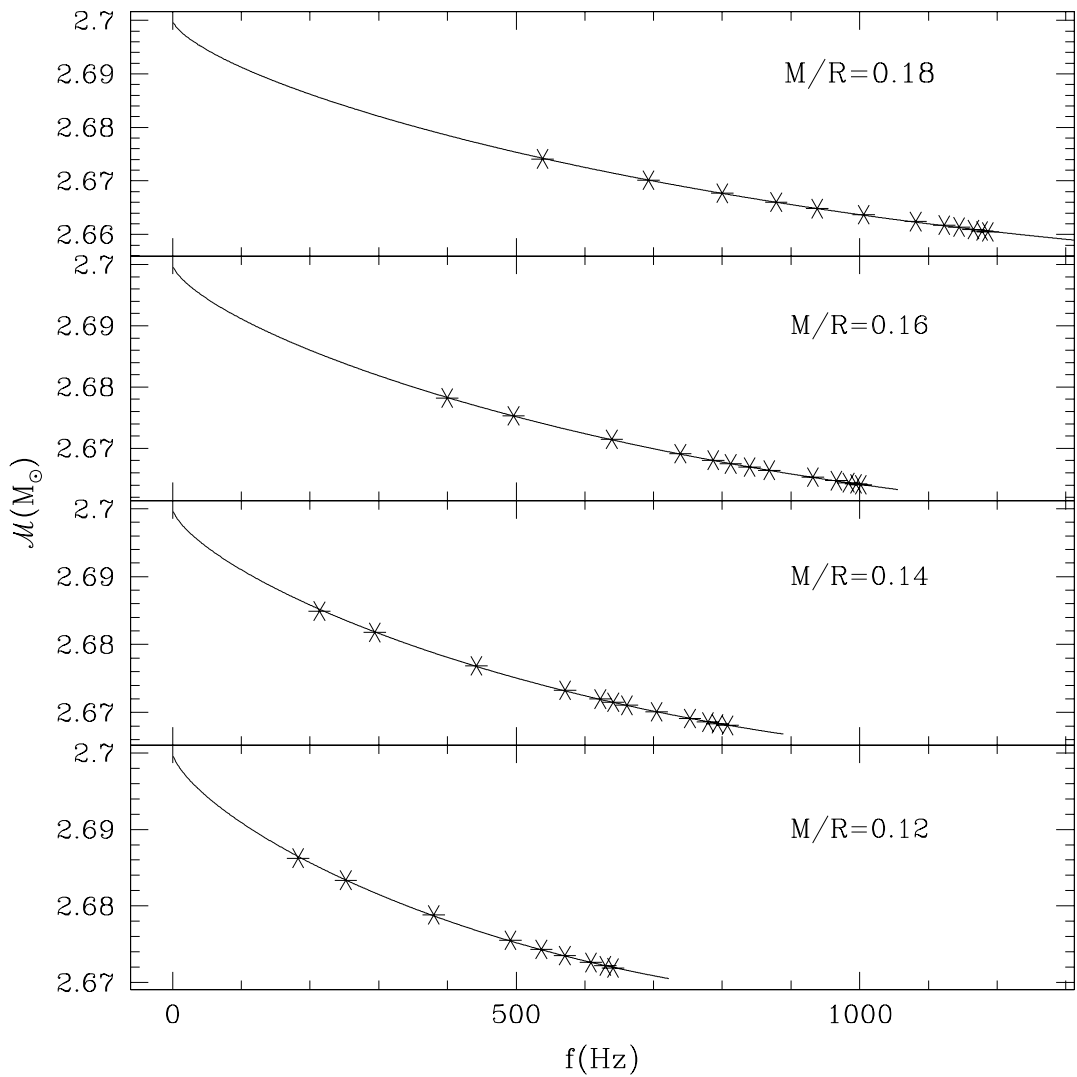}
\caption{ADM mass (total mass-energy) of a binary NS system
as a function of GW frequency (twice the orbital frequency), 
computed along each of our 4 irrotational equilibrium sequences.  
From bottom to top, the sequences correspond to NS with
compactness $M/R=\,0.12, 0.14, 0.16$, and $0.18$.
All models assume a polytropic EOS with $\Gamma=2$ and a NS mass of 
$1.35\, M_\odot$ for both components.  The asterisks indicate the 
individual equilibrium
configurations calculated along each sequence, while the lines show
our best fit using Eq.~\protect\ref{eq:fit} and the values of Table~1.}
\label{fig:plotmom}
\end{figure}

For each NS model, about 12 binary equilibrium configurations
are computed with decreasing
separations, until a cusp develops on the NS surface.
Equilibrium configurations for smaller separations do not exist.
To each sequence we fit a curve of the form
\begin{equation}
{\cal M}(f)=2.7 M_\odot - k_{\rm N} f^{2/3} + k_1 f + k_2 f^2
\label{eq:fit}
\end{equation}
to represent the variation of total mass-energy as a function of GW 
frequency.
The first term gives the total gravitational mass of the system
at infinite separation, while the second term represents the Newtonian
point-mass behavior, with 
$k_{\rm N}=2^{-4/3}\pi^{2/3}G^{2/3}M^{5/3}=4.056\times 10^{-4}\, M_\odot\,
{\rm Hz}^{-2/3}$.   
The term $\propto f$ was introduced heuristically to represent the lowest-order
post-Newtonian and finite-size corrections to the point-mass behavior at
intermediate frequencies. The term $\propto f^2$ represents the tidal
interaction energy, which causes the equilibrium energy curve to
flatten at high frequencies.
Our best (least-squares) fit values of $k_1$ and $k_2$
for each sequence are listed in Table~1 and the results are illustrated in
Fig.~\ref{fig:plotmom}. The asterisks show the data points along each
sequence, with a typical error between the data points and the fit of
$\delta {\cal M}\sim10^{-4}\,M_\odot$.  We find in all cases that $k_2$ is positive, as
we would expect: tidal deformations and relativistic gravitational
effects increase the equilibrium energy
\cite{LRS3,LomRS}. We
note that none of the equilibrium curves shows evidence of an energy
minimum, which would have implied the onset of dynamical
instability \cite{Lai,TaniII,LRS3,LRS2}.

\begin{figure}
\centering \leavevmode \epsfxsize=6in \epsfbox{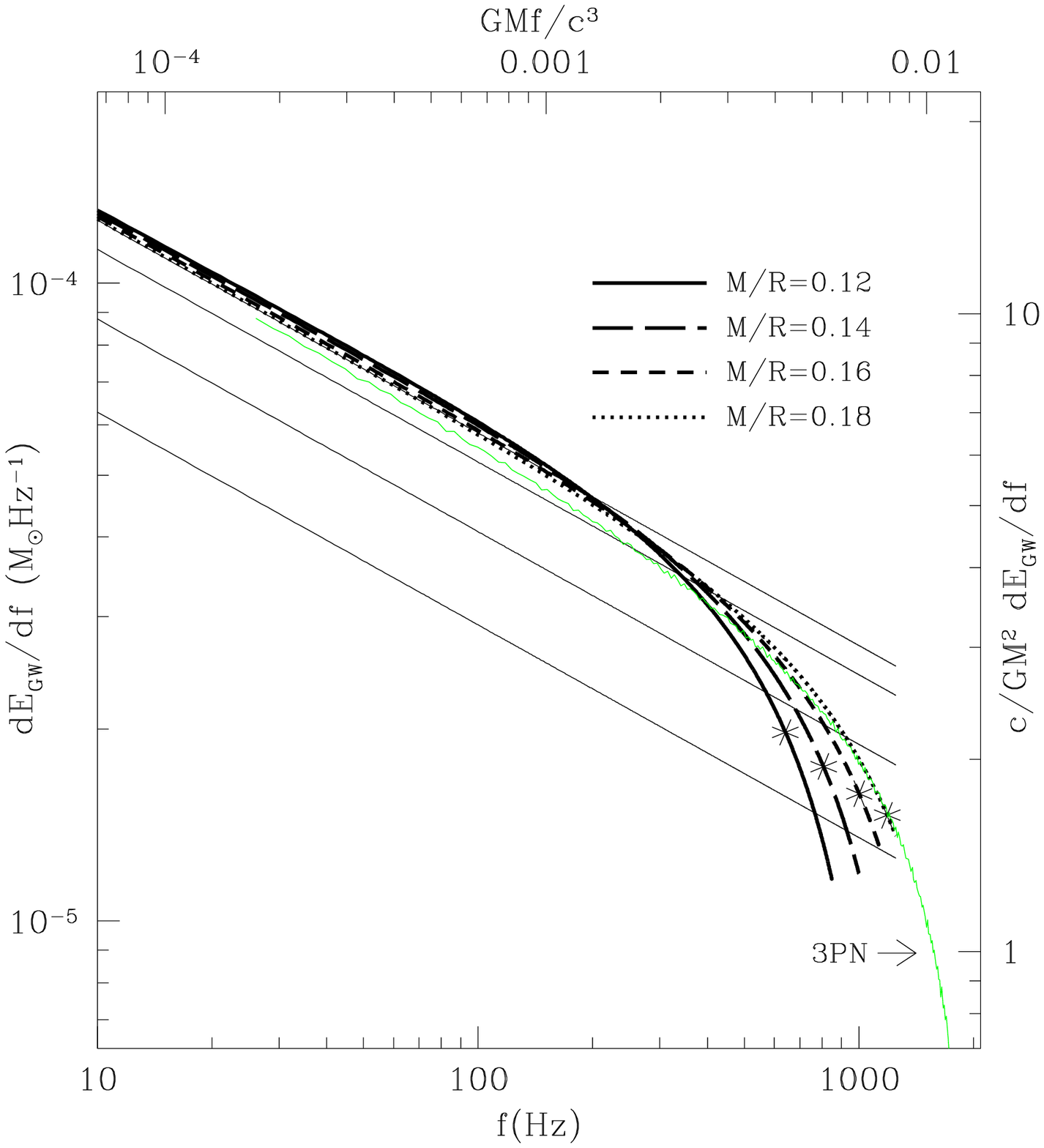}
\caption{Energy spectrum $dE_{\rm GW}/df$ 
of GW emission emitted along
each of the 4 sequences of Fig.~\protect\ref{fig:plotmom} in the 
quasi-equilibrium approximation.  Also shown is an irrotational
3PN point mass binary from Ref.~\protect\cite{Damour}, which closely
tracks our most compact model.  Asterisks indicate the
terminal point along each sequence, where a cusp develops.  
The slanted straight lines show,
from top to bottom, the point-mass Newtonian energy spectrum
($\propto f^{-1/3}$) multiplied by 1.0, 0.9, 0.75, and 0.5.
The last three values are used to define characteristic break
frequencies $f_{10}$, $f_{25}$, and $f_{50}$, where the energy
spectrum has dropped by the corresponding fraction.  The
units on the right and top axes show the corresponding dimensionless
quantities, with the mass dependence scaled away.} 
\label{fig:dedf}
\end{figure}

Computing the GW energy spectrum for each model now only requires
differentiating the fitted curves with respect to frequency. 
The results are shown in Fig.~\ref{fig:dedf}. In each case, we see
a characteristic frequency range where the spectrum plunges rapidly
below the extrapolation of the low-frequency result. 
This corresponds to the flattening of the
energy curves in response to the growing tidal interaction and PN effects.
Also shown is the energy spectrum of a 3PN, irrotational, point-mass
binary, computed according to the results found in Ref.~\cite{Damour},
which closely tracks our most compact model, indicating that the
differences we see in the energy spectra result from finite-size
effects associated with the NS radius.
To quantify the deviations from the Newtonian case, 
we define a set of break frequencies,
at which the energy spectrum has dropped by some factor below the
point-mass result.
The values we find for $f_{10},~f_{25}$, and $f_{50}$, where
$dE_{\rm GW}/df$ has dropped by $10\%,~25\%$, and $50\%$,
respectively,  are listed in the last three columns of Table~1. 
We see that all these characteristic
frequencies lie within the frequency range accessible by LIGO-type
detectors,
with perhaps the exception of $f_{50}$ for the more compact sequences.
Note that the calculation of $f_{50}$ values requires extrapolating 
the equilibrium energy curves slightly beyond the last equilibrium model 
(where a cusp develops), and may therefore be less reliable.
However, we find that $f_{50}$ has a particularly steep,
quasi-linear dependence on the NS compactness, given by 
$f_{50}\simeq [10^4 (M/R)-460]\,{\rm Hz}$ within the range of
NS radii we considered. For comparison,
$f_{25}$ values can be determined safely within the quasi-equilibrium
approximation, and the sensitivity on compactness is only slightly
reduced, with $f_{25}\simeq [5000(M/R)-85]\,{\rm Hz}$.

The best definition of the break frequency will be a tradeoff 
between higher signal-to-noise ratio at lower frequencies \cite{LIGOnote},
and greater ease of discrimination between different EOS at the
higher frequencies. In addition, the quasi-equilibrium approximation
is expected to be most accurate at lower frequencies, where the inspiral
rate is slower. However, we doubt that this could become a major issue:
if we adopt, for simplicity, the point-mass formula for the GW luminosity, 
and compute the corresponding radial infall velocity along our
equilibrium sequences, we find that $v_r$ never exceeds $5\%$ of the 
orbital velocity, even at the point where we define $f_{50}$ (the
corresponding fraction at the point where we define $f_{25}$ is about
2\%) \cite{NOTEvr}.
Ultimately, the optimum choice should be
determined by data analysts, 
taking into account the accuracy with which
the break frequencies can be extracted using matched filtering or other
techniques \cite{CF,Match}.
Preliminary studies of this problem
have already been performed for both broad-band and
narrow-band interferometer configurations \cite{Vallis,Hughes}.
Defining a precise break frequency may not even be necessary.
Instead, the GW inspiral templates could be terminated
at high frequency in a manner that reproduces the energy spectrum 
given by a simple analytic form, such as our Eq.~\ref{eq:fit}.
The free parameters $k_1$ and $k_2$ could then be measured experimentally
and compared directly to the predictions of various realistic nuclear
EOS used in computing binary equilibrium sequences. 
In future work, we plan
to compute such sequences, and the corresponding energy curves, for
a wide variety of published, realistic NS EOS.

We are grateful to Kip Thorne for originally pointing out the importance of
the break frequencies in GW spectra. This work was supported by NSF Grants
PHY-0070918, PHY-0121420, and PHY-0133425.

\newpage
\begin{table}
\caption{Properties of the quasi-equilibrium sequences.
Here $M/R$ (with $G=c=1$) is the compactness of an isolated NS seen by an observer at
infinity, $R$ is the circumferential radius for an ADM mass of
$M=1.35\, M_{\odot}$, $f_{\rm c}$ is the GW frequency at the final point
of each sequence (cusp), $k_1$ and $k_2$ are the best fit parameters
in Eq.~\protect\ref{eq:fit}, and $f_{10}$, $f_{25}$, and
$f_{50}$ are the break frequencies at which the GW energy spectrum has dropped,
respectively, by
$10\%$, $25\%$, and $50\%$ below the point-mass energy spectrum.}
\label{table:table1} 

\begin{tabular}{c|ccccccc}
$M/R$ & $R\,$(km) & $f_{\rm c}\,$(Hz) & $k_1$ & $k_2$ & $f_{10}\,$(Hz) & $f_{25}\,$(Hz) &
$f_{50}\,$(Hz) \\
\colrule\colrule
0.12 & 16.6 & 641 & -4.939E-6 & 1.290E-8 & 342 & 518 & 764 \\
0.14 & 14.2 & 807 & -3.363E-6 & 9.244E-9 & 383 & 612 & 931 \\
0.16 & 12.4 & 1002 & -1.806E-6 & 6.490E-9 & 418 & 720 & 1137 \\
0.18 & 11.1 & 1187 & -5.834E-7 & 4.835E-9 & 431 & 810 & 1331 \\
\end{tabular}
\end{table}

\newpage

\newpage


\begin{thebibliography}{DUM}
\bibitem{NOTEvr} Even when finite-size effects are taken into account in full
GR calculations, the radial infall velocity still does not
exceed about $15\%$ of the orbital velocity at any point along
the equilibrium sequence for an assumed compactness of $M/R=0.19$, and for
a compactness $M/R=0.14$, the ratio never exceeds $5\%$; 
see Fig.~15 of Ref.~\cite{SU}.
\end{thebibliography}
\end{document}